# The role of Trees of Fragmenting Granules (TFG) in the formation of the solar supergranular pattern from Hinode observations

*Jean-Marie Malherbe (1) and Thierry Roudier (2)*

(1) Emeritus astronomer, Observatoire de Paris, PSL Research University, LIRA, France
Email: Jean-Marie.Malherbe@obspm.fr; ORCID: https://orcid.org/0000-0002-4180-3729

(2) Honorary research director, Observatoire Midi Pyrénées, IRAP, France
Email: thierry.roudier@irap.omp.eu

11 March 2025

**ABSTRACT**

We present in this paper an exceptional scientific dataset allowing to investigate the structure and evolution of the interior of solar supergranulation cells. Trees of Fragmenting Granules (TFG) and associated flows were evidenced using Local Correlation Tracking techniques (LCT) from a 24 H duration sequence of Hinode (JAXA/NASA) observations. The treatment of the dataset exhibits the evolution of the TFG and shows that their mutual interactions are able to build horizontal flows with longer lifetime than granules (1 to 2 hours) over a scale of 10 arcsec (the mesogranulation). These flows act on the diffusion of the intranetwork magnetic elements and also on the location and shape of the network. Hence, the TFG appear as one of the major elements involved in supergranular formation and evolution.

**INTRODUCTION**

Hinode is a Japanese mission developed and launched by ISAS/JAXA, with NAOJ as domestic partner and NASA and STFC (UK) as international partners. It is operated by these agencies in co-operation with ESA and NSC (Norway).

We obtained an exceptional sequence of observations on 29-30 August 2007, lasting 24 H, with the 50 cm diameter Solar Optical Telescope (SOT). We used the Broad band Filter Imager (BFI) in the 4500 Å blue continuum and in the 4305 Å G-band, together with the Narrow band Filter Imager (NFI, a Lyot filter of 60 mÅ bandwith) and polarimeter running the Stokes I-V mode in FeI 6302 Å line. The pixel size was 0.11 arcsec for the BFI, 0.16 arcsec for the NFI, and the time step 50.2 s (1720 steps). The selected FOV (60 arcsec x 62 arcsec) was at disk centre and included a superganule, as revealed by the quasi closed magnetic network delineating its boundaries.

**DATA PROCESSING**

We did not used the G-band (available in the dataset) because of bright points perturbing the segmentation of the granular structure. We preferred to use the 4500 Å continuum for that purpose. The 24 H duration sequence (with rotation compensation activated) was then realigned and filtered from 5-minutes photospheric oscillations via a subsonic filter in the Fourier (k, ω) diagram. Hence, two data treatments allowed to derive families of granules and horizontal velocity flows. A third step consisted in introducing and following corks transported by the flow. Of course, other approaches could be undertaken using the accompanying dataset.

TFG or families were derived using the segmentation and labelling algorithm developed by Roudier *et al* (2003). Exploding granules form a set of evolving TFG as shown by Roudier *et al* (2016) and Malherbe *et al* (2018). Each granule of a family is identified by a common integer number, which allows to see the formation, follow the development and decay of each family. Horizontal velocities were computed using the LCT technique proposed by November & Simon (1988). It uses a temporal window of 30 minutes and spatial window of 1.5 arcsec, so that it evidences only the average and large scale velocity field, typically at the size of the mesogranulation (10 arcsec). Corks were a useful tool to follow the horizontal motions, so that 9000 corks were introduced at time t = 0 in the flow, in order to identify (or not) structures possibly formed by their final position.

**MAIN RESULTS**

Figure 1 shows the action of the mesogranular flow computed from the LCT on the formation of the magnetic network. The dataset reveals expanding motions that transport outwards the intranetwork magnetic elements, forming supergranular boundaries. Such motions are the result of the formation of large TFG which form and develop inside the supergranular cell (details in Roudier *et al*, 2003, 2016). As there are many concurrent and adjacent cells on the Sun running the same process, concentrations of magnetic fields thus appear at the borders of the cells, this is the chromospheric network. Figure 2 shows the trajectories and the final positions of corks pushed by the flow generated by the evolving TFG (9000 corks); their final location delineates the supergranular boundaries.

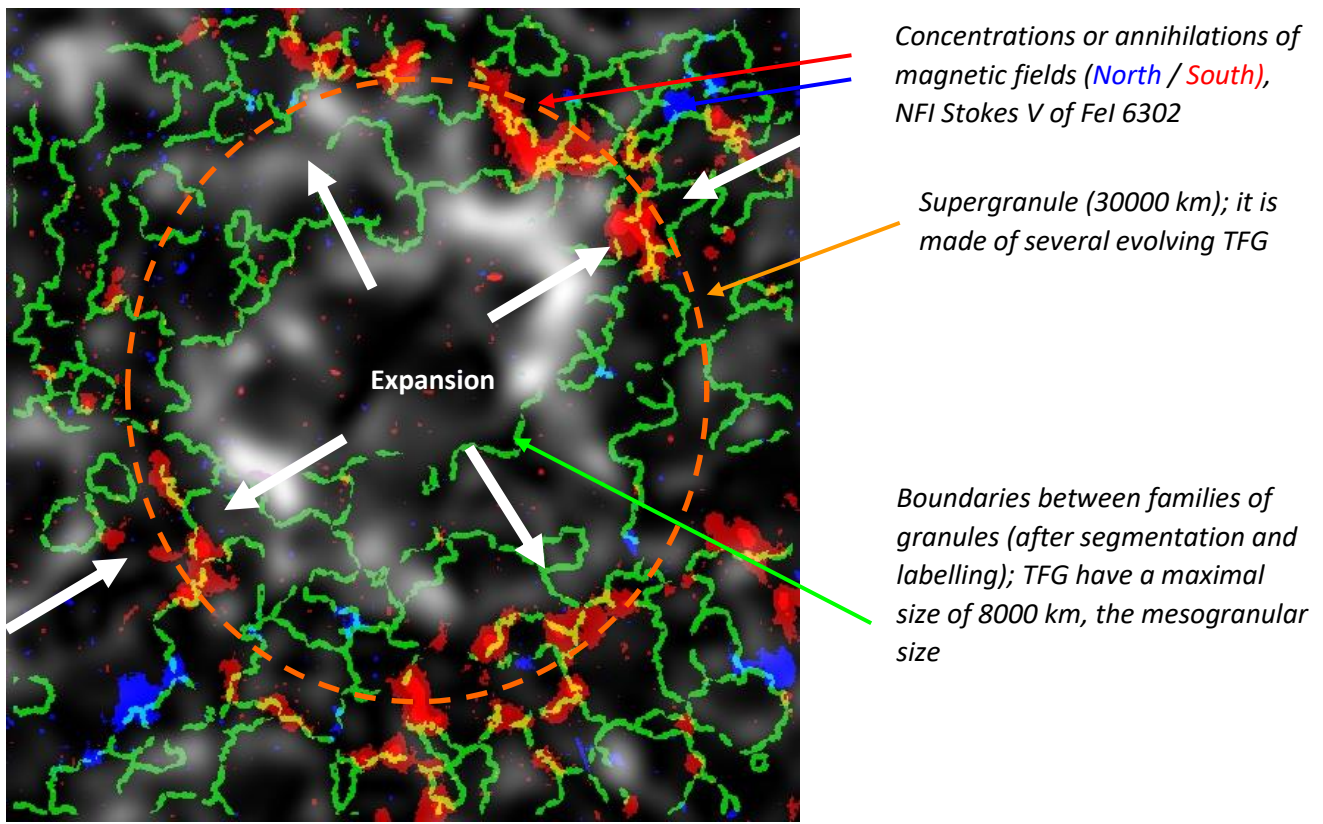

*Concentrations or annihilations of magnetic fields (North / South), NFI Stokes V of FeI 6302*

*Supergranule (30000 km); it is made of several evolving TFG*

*Boundaries between families of granules (after segmentation and labelling); TFG have a maximal size of 8000 km, the mesogranular size*

*Figure 1 : a dynamic model of a supergranule (1' x 1' FOV). The background Image shows horizontal velocities from Local Correlation Tracking (LCT) of the blue continuum, $V_h = (V_x^2 + V_y^2)^{½}$ in grey levels.*

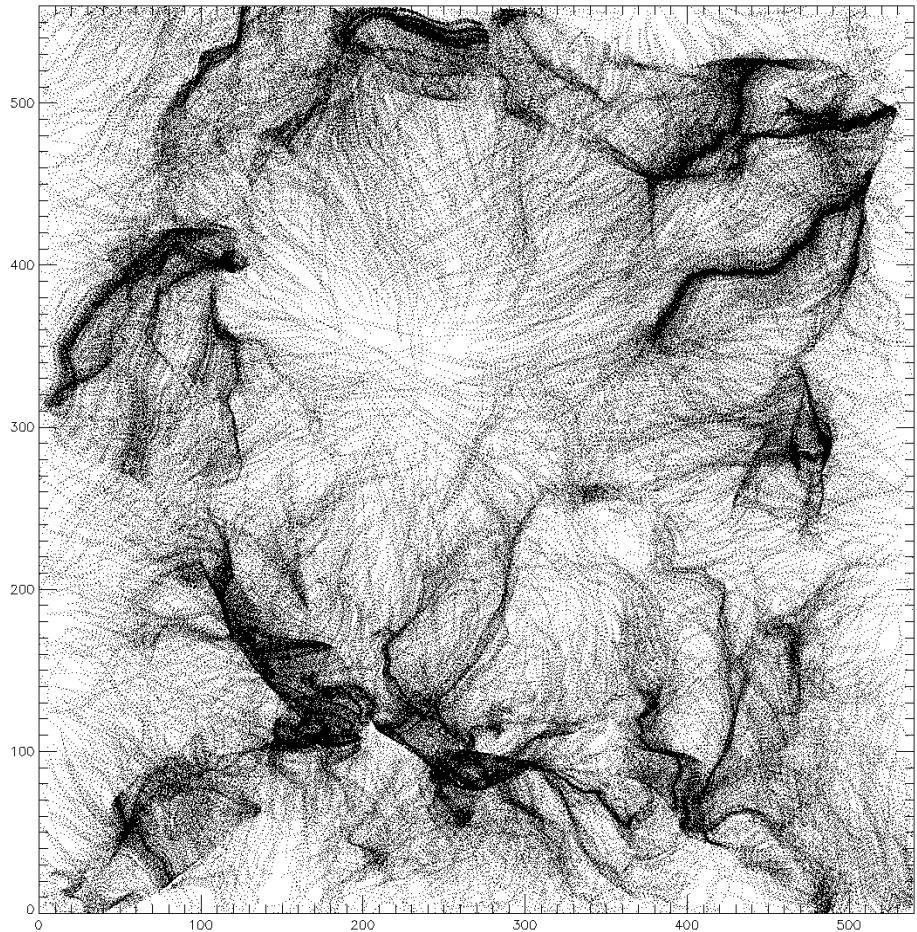

*Figure 2 : the trajectories of corks during the 24 H sequence. They were uniformly distributed at t = 0, the trajectory of 9000 corks is displayed. They finally delineate the supergranule.*

**ASSOCIATED MOVIES (MPEG 4 format)**

**ZIP file "Movies.zip" at FigShare.com : https://figshare.com/s/259fe9f7ebbd99494ee4**

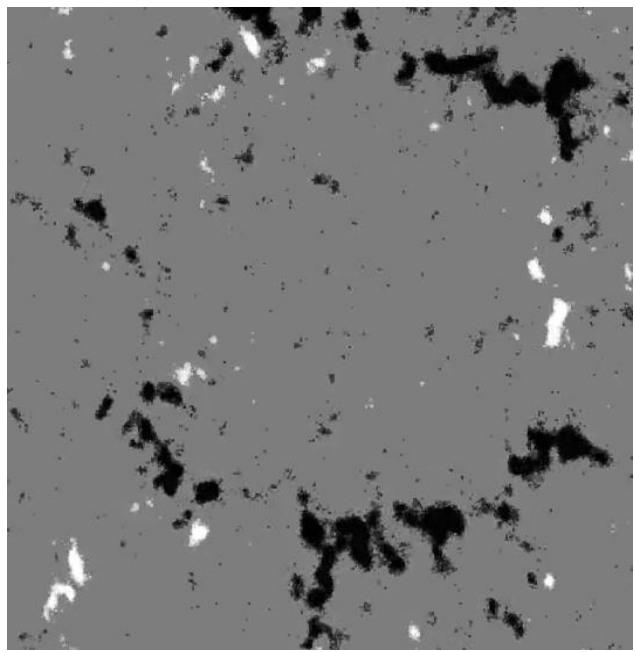

*Movie 1 :*

*LOS magnetic field (SOT/NFI FeI 6302), polarities in black/white, time step 50 s, 24 H duration, FOV 540 x 560 pixels of 0.11 arcsec (1 x1 arcmin²)*

*Movie 2 : Trees of Fragmenting Granules (or families), LOS magnetic field (polarities in yellow/white), time step 50 s, 24 H duration, FOV 540 x 560 pixels of 0.11 arcsec (1 x1 arcmin²). Granules belonging to the same TFG use the same colour.*

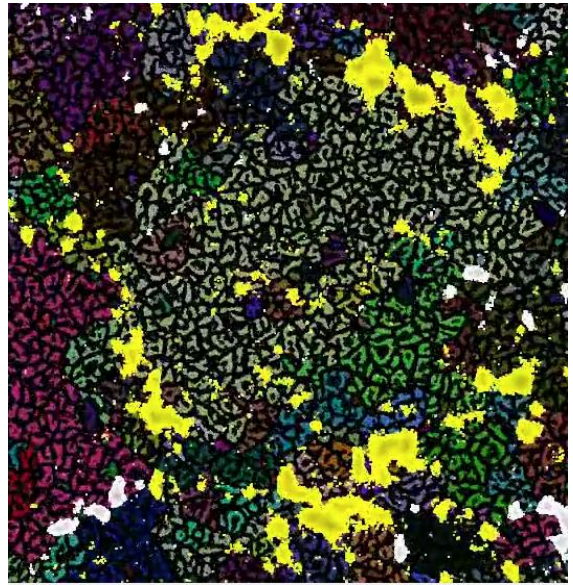

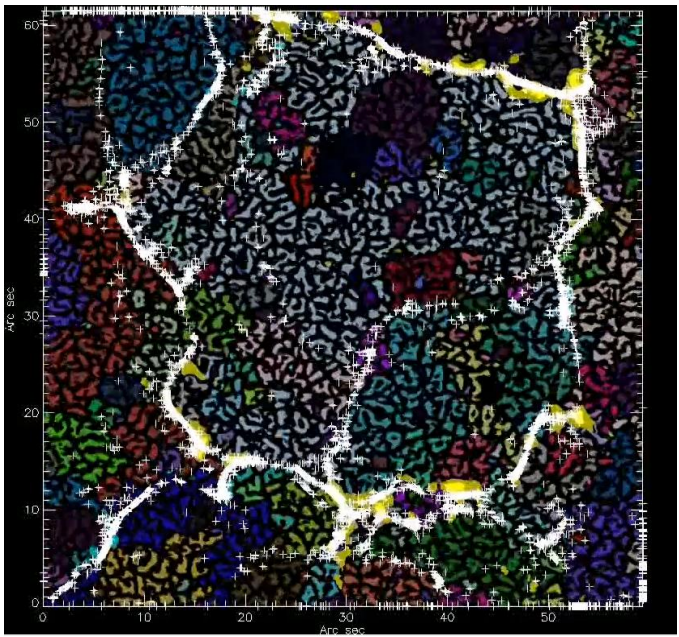

*Movie 3 : Trees of Fragmenting Granules (or families), LOS magnetic field (polarities in yellow/white), and evolution of corks positions (white crosses), time step 180 s, 24 H duration, FOV 540 x 560 pixels of 0.11 arcsec (1 x1 arcmin²). Granules belonging to the same TFG have the same colour.*

*Movie 4 : LOS magnetic field (polarities in red/blue), and evolution of the horizontal velocity $V_h = (V_x^2 + V_y^2)^{1/2}$ derived from LCT, time step 180 s, 24 H duration, FOV 540 x 560 pixels of 0.11 arcsec (1 x1 arcmin²). High velocities are white, while null velocities are black.*

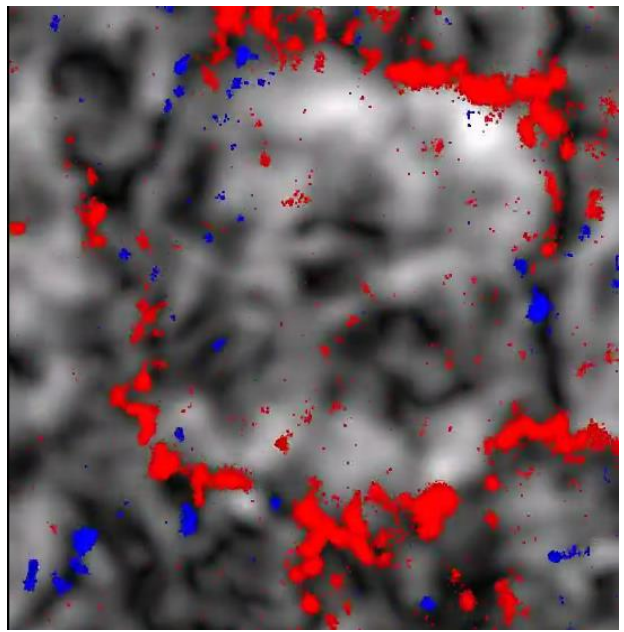

**ASSOCIATED SCIENTIFIC DATASET (FITS FORMAT)**

On line at FigShare.com : https://figshare.com/s/259fe9f7ebbd99494ee4

DOI : 10.6084/m9.figshare.28559339

ZIP file : **Hinode2007.zip**

FITS data were compressed by FPACK; please use the FUNPACK tool for decompression.

Sequence duration : 24 H (50 s x 1720 steps)

Hinode SOT/NFI, LOS magnetic field (Gauss), time step 50 s : **b.fits.fz**

Hinode SOT/BFI, Blue continuum 4500 Å, time step = 50 s : **bluecont.fits.fz**

Hinode SOT/BFI, G-band 4307 Å, time step = 300 s : **G-band.fits.fz**

Trees of Fragmenting Granules derived from NFI 4500 Å, time step = 50 s : **families.fits.fz** (exploding granules form trees of fragmenting granules; granules of the same tree or family are labelled by a common long integer number)

Also derived from BFI 4500 Å :

Mean horizontal velocity Vx (in m/s), from LCT using 1800 s and 1.5 arcsec windows, time step = 1800 s : **vx.fits.fz**

Mean horizontal velocity Vy (in m/s), from LCT, time step = 1800 s : **vy.fits.fz**

X location of corks (x 1000, in pixels, 1 pixel = 0.11 arcsec), time step = 180 s : **xcorks.fits.fz**

Y location of corks (x 1000, in pixels, 1 pixel = 0.11 arcsec), time step = 180 s : **ycorks.fits.fz**

NP number of successive positions of each cork : **nposxycorks.fits.fz**

There are nc = 9000 corcks ; for cork number n, the successive positions as a function of time t (step 180 s) are : [x(n,0:NP(n)-1), y(n,0:NP(n)-1)]; the lifetime of cork n is hence NP(n) x 180 s.